\documentstyle[psfig,conf_iap,]{article}
\begin{document}
\heading{%
%Begin Heading
%
The N/O abundance ratio in the lowest-metallicity blue compact
dwarf galaxies \\
%
%End Heading
} 
\par\medskip\noindent
\author{%
%Begin Author names
Yuri Izotov$^{1}$, Frederic Chaffee$^{2}$, Craig Foltz$^{3}$, 
Klaus Fricke$^{4}$, Richard Green$^{5}$, Natalia Guseva$^{1}$, 
Kai Noeske$^{4}$, Polychronis Papaderos$^{4}$, Trinh Thuan$^{6}$
%End Author names
}
\address{%
%First address
Main Astronomical Observatory, 27 Zabolotnoho str., 03680 Kyiv, Ukraine.
}
\address{%
%Second address
W. M. Keck Observatory, 65-1120 Mamalahoa Hwy., Kamuela, HI 96743, USA.
}
\address{%
%Third address
MMT Observatory, University of Arizona, Tucson, AZ 85721, USA.
}
\address{%
%Forth address
Universit\"ats--Sternwarte, Geismarlandstra\ss e 11, D--37083 G\"ottingen, 
Germany.}
\address{%
%Fifth address
National Optical Astronomy Observatories, Tucson, AZ 85726, USA.
}
\address{%
%Sixth address
Astronomy Department, University of Virginia, Charlottesville, VA 22903, USA.
}

\begin{abstract}
The results of the N/O abundance determination in a sample
of low-metallicity blue compact dwarf (BCD) galaxies based on new
spectroscopic observations with large telescopes (Keck, VLT, MMT, 4m KPNO)
are presented. We show that the N/O abundance ratio is constant at lowest metallicities $\leq$ $Z_\odot$/20, confirming previous findings and strongly
supporting the origin of nitrogen as a primary element.
\end{abstract}
\section{Sample}
The sample consists of $\sim$ 100 BCDs with a heavy element 
abundances in the range $Z_\odot$/50 -- $Z_\odot$/4, selected from the
First and Second Byurakan, Michigan, Hamburg, Tololo and Case surveys.
All most metal-deficient BCDs with $Z$ $\leq$ $Z_\odot$/20 are included.
High signal-to-noise ratio spectroscopic observations were made during 1993 -- 2000 with Keck, VLT, MMT and 4m KPNO telescopes. In total, the results
of 119 independent observations of H {\sc ii} regions are presented in this
contribution.
\section{Results}

The dependence of the [N/O] vs [O/H] abundance ratios is shown in Fig. 1,
where [N/O] = log(N/O) -- log(N/O)$_\odot$ and [O/H] = log(O/H) -- 
log(O/H)$_\odot$.
The open circles show data from \cite{IT99} and solid circles are new data.
We conclude from Fig. 1 that

\begin{itemize} 

\item the N/O abundance ratio is constant in BCDs with [O/H] $<$ --1 and
favors the primary production of nitrogen by intermediate-mass and/or massive
stars. Secondary nitrogen production is important for higher metallicities;

\item the very small dispersion of the N/O abundance ratio in BCDs with 
[O/H] $<$ --1.4 favors the primary production of nitrogen in massive stars;

\item the apparent difference of the N/O abundance ratios in BCDs and damped 
Ly$\alpha$ absorption (DLA) systems (e.g. \cite{ISC00,IT99}) 
could be caused
by the different enrichment scenarios or unaccounted ionization effects in 
DLAs.

\end{itemize}

\begin{figure}
\centerline{\vbox{
\psfig{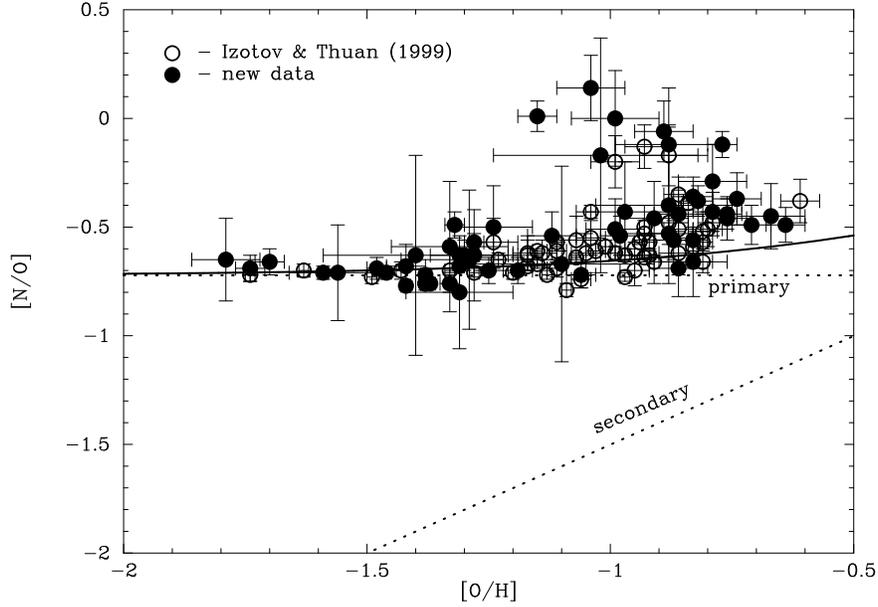}
}}
\caption[]{[N/O] vs [O/H] diagram for BCDs. Open circles
are from \cite{IT99}, 
filled circles are new data. Dotted lines are diagrams for primary and
secondary productions of nitrogen, while the solid line shows the dependence 
[N/O] vs [O/H] when both primary and secondary productions of nitrogen are 
taken into account.}
\end{figure}

\acknowledgements{Y.I. and N.G. have been partially supported by INTAS 
97-0033 and Swiss SCOPE 7UKPJ62178 grants. 
Those authors, P.P. and K.F. acknowledge support by the Volkswagen 
Foundation under grant No. I/72919. Y.I. and T.T. have been partially 
supported by NSF grant AST-9616863. Research by P.P. and K.F. has been 
supported by the
Deutsches Zentrum f\"{u}r Luft-- und Raumfahrt e.V. (DLR) under
grant 50\ OR\  9907\ 7. C.F. acknowledges the support of the NSF under grant
AST-9803072, and K.N. of the Deutsche Forschungsgemeinschaft (DFG) grant 
FR325/50-1.}

\vfill
\end{document}